\documentclass[twocolumn,aps,pra,longbibliography,superscriptaddress]{revtex4}
\usepackage{graphicx}
\usepackage{amsmath}
\usepackage{amssymb}
\usepackage{braket}
\usepackage{physics}
\usepackage[colorlinks=true,linkcolor=blue,citecolor=blue,urlcolor=blue]{hyperref}

\begin{document}

\title{Tunable Confinement-Deconfinement Transition in an Ultracold Atom Quantum Simulator }

\author{Yanting Cheng}
\affiliation{Institute of Theoretical Physics and Department of Physics, University of Science and Technology Beijing, Beijing 100083, China}
\author{Shang Liu}
\affiliation{Kavli Institute for Theoretical Physics, University of California,
	Santa Barbara, California 93106, USA}
\author{Wei Zheng}
\affiliation{Hefei National Laboratory for Physical Sciences at the Microscale and
Department of Modern Physics, University of Science and Technology of China,
Hefei 230026, China}
\affiliation{CAS Center for Excellence in Quantum Information and Quantum Physics,
University of Science and Technology of China, Hefei 230026, China}

\author{Pengfei Zhang}
\affiliation{Institute for Quantum Information and Matter and Walter Burke Institute for Theoretical Physics, California Institute of Technology, Pasadena, California 91125, USA}
\author{Hui Zhai}
\email{hzhai@tsinghua.edu.cn}
\affiliation{Institute for Advanced Study, Tsinghua University, Beijing 100084, China}
\date{\today}
\begin{abstract}

The one-dimensional lattice Schwinger model has recently been realized by using bosons in optical lattices. This model contains both confinement and deconfinement phases, whose phase diagram is controlled by the mass of the matter field and the topological angle. Since varying the mass of matter field is straightforward experimentally, we propose how to tune the topological angle, allowing accessing the entire phase diagram. We propose that direct experimental evidence of confinement and deconfinement can be obtained by measuring whether a physical charge is localized around a fixed gauge charge to screen it. We also discuss the PXP model realized in the Rydberg atoms array, which is equivalent to the lattice Schwinger model when all local gauge charges are fixed as zero. Although the gauge charges are fixed, we can alternatively probe the confinement and the deconfinement in the PXP model by studying the relative motion of a pair of a physical charge and an anti-charge. Our scheme can be directly implemented in these two relevant experimental platforms of ultracold atom quantum simulators. 

\end{abstract}
\maketitle

\section{Introduction}

Confinement-deconfinement (conf-deconf) transition is an important phenomenon in nature, ranging from quark confinement in high-energy physics  \cite{quark_confinement} to vortex deconfinement driven Kosterlitz-Thouless transition in condensed matter and ultracold atom systems \cite{KT}. Recently, an experiment has simulated the one-dimensional lattice Schwinger model \cite{Schwinger1,Schwinger2,Coleman,QLM}, which is a $U(1)$ lattice gauge theory (LGT) \cite{Kogut1,Kogut2}, using ultracold bosons in optical lattices \cite{USTC}. The progress along this direction has also drawn considerable attentions theoretically \cite{LGTtheory1,LGTtheory2,LGTtheory3,Peter2012,LGTtheory4,Peter2013,Peter2014,Peter2016,LGTtheory5,LGTtheory6,LGTtheory7,LGTtheory8,LGTheory12,LGTtheory9,LGTtheory10,LGTtheory11,LGTtheory13}. The Schwinger model has a parameter called topological angle $\theta$, and the conf-deconf transition exists in this model only when $\theta=\pi$ \cite{David}. In this experiment, it is precisely the model with $\theta=\pi$ that has been realized, and the signal of a phase transition has been observed \cite{USTC}. This offers an opportunity to study the conf-deconf phase transition in a highly controllable setting. 

In this work, we will discuss two important issues for further experimental studies of this lattice gauge model. The first issue is how to provide direct experimental evidence for confinement and deconfinement of two phases before and after the transition. The second issue is how to tune the topological angle experimentally, and therefore tune the conf-deconf transition. We will present our theoretical proposals on both issues. 

Another recent experimental development in ultracold atom physics is the Rydberg atom array quantum simulator~\cite{Rydberg2016,Rydberg2017,Rydberg2019,Rydberg2019Lukin,Rydberg2021,Rydberg2021Lukin,Rydberg2021Lukin2}. Because of the physics of the Rydberg blockade, this system is described by the so-called PXP Hamiltonian~\cite{PXP}. This Rydberg atom quantum simulator can simulate quantum many-body dynamics with high fidelity and explore exotic ground state quantum phases \cite{PXPtheory1,PXPtheory2,PXPtheory3,PXPtheory4,PXPtheory5,PXPtheory10,PXPtheory6,PXPtheory7,PXPtheory11,PXPtheory8,PXPtheory12,PXPtheory9,PXPtheory13,PXPtheory14,Shang,Shang2,Cheng,PXPspin-liquid,PXPtheory15}. Here we will also discuss the connection between the $U(1)$ LGT and the PXP Hamiltonian \cite{LGTtheory8} and discuss the manifestation of the conf-deconf transition in the PXP model. 

This paper is organized as follows. We first review the experimental setting in Sec II. Then, we will discuss how to tune the topological angle $\theta$ in Sec III and present the protocol for detecting confinement or deconfinement in Sec IV. In Sec V, we will bring out the connection between this LGT and the PXP model, and we will show that the conf-deconf physics can also be studied in the PXP model. We summarize the results in Sec VI. 

\begin{figure}[t]
\centering
\includegraphics[width=1\linewidth]{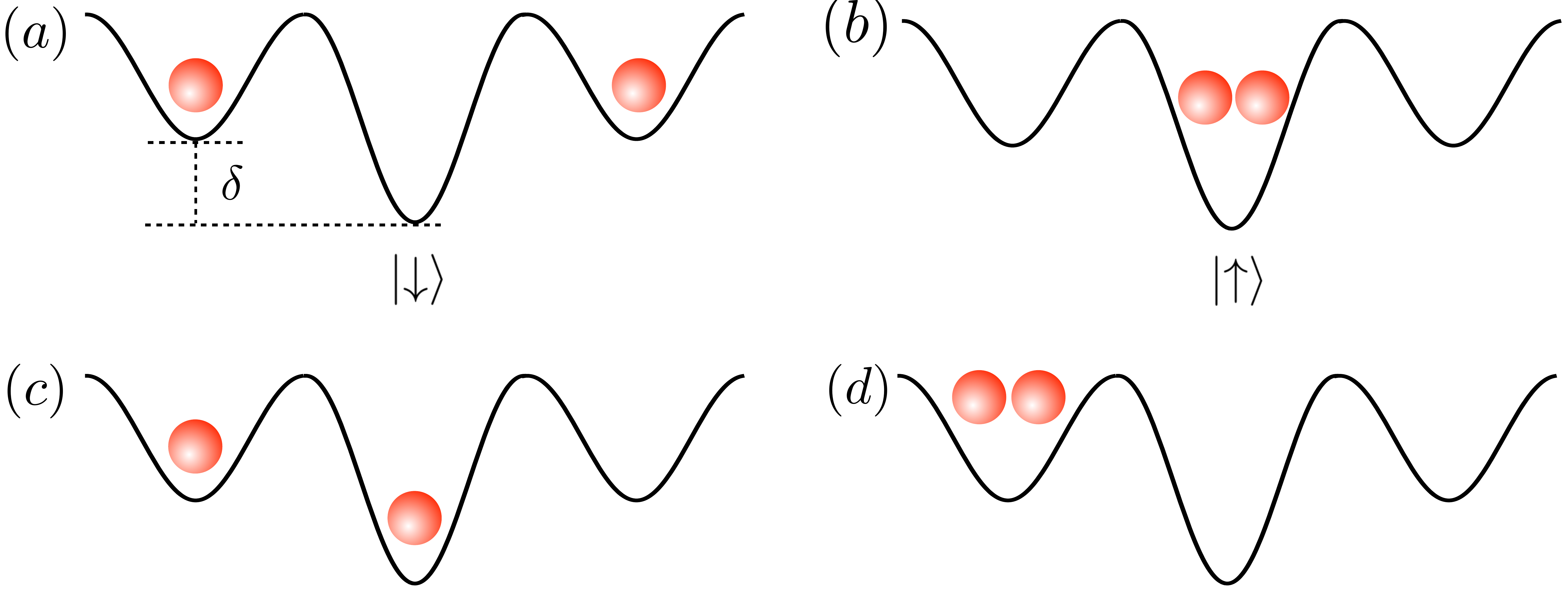}
\caption{Schematic of different states of a three-site building block. When $U\approx2\delta$ and $J\rightarrow 0$, the energies of (a) and (b) are nearly degenerate, and are off-resonance with the energies of (c) and (d).}
\label{projection}
\end{figure}

\section{A Brief Review of the Experiment}

The experiment of Ref. \cite{USTC} uses bosons in a one-dimensional optical lattice with a staggered potential, and the Hamiltonian is the Bose-Hubbard model that reads
\begin{equation}
\hat{H}_{\textrm{BHM}}=\sum_j[-J(\hat{b}_j^\dagger\hat{b}_{j+1}+h.c.)+\frac{U}{2}\hat{n}_j(\hat{n}_j-1)+\epsilon_j\hat{n}_j],
\label{BHMH}
\end{equation} 
where $\epsilon_j=(-1)^j\delta/2$ is the staggered potential, and $\delta$ sets the on-site energy difference between the even and the odd sites. $\hat{b}_j^\dagger$ and $\hat{b}_j$ are the creation and annihilation operators. $\hat{n}_j=\hat{b}_j^\dagger\hat{b}_j$, $J$ denotes the tunneling strength between neighboring sites, and $U$ denotes the on-site interaction strength. In practice, there is also a gradient potential that suppresses long-range hopping. Since we do not explicitly include the long-range hopping in our model, we will ignore this gradient potential term as well.  

This experiment works in the parameter range that $\delta\gg J$, $U\gg J$ and $U\approx2\delta$ \cite{USTC}. We show four different cases for two-atom states in Fig. \ref{projection}. In the limit $J\rightarrow 0$ and $U=2\delta$, the energy is $\delta$ when two even sites are singly occupied (Fig. \ref{projection}(a)), and the energy is $U-\delta$ when an odd site is doubly occupied (Fig. \ref{projection}(b)). Thus, the energies of these two states are nearly degenerate. However, for states as shown in Fig. \ref{projection}(c) and (d), the energies are $0$ and $U+\delta$, respectively, and these two energies are offset by $\delta$ and 2$\delta$, compared with states shown in Fig. \ref{projection}(a) and (b). Hence, when $\delta \gg J$, we can project out both the doubly occupied even sites or the singly occupied odd sites. That is to say, for odd sites, we only retain the vacuum state and the doubly occupied state. We view these two cases as a spin-$1/2$, with the vacuum state denoted by $\ket{\downarrow}$ and the doubly occupied state denoted by $\ket{\uparrow}$. These spin-$1/2$ states are viewed as the degree-of-freedoms of the gauge field.

Now considering a small but finite hopping amplitude $J$, the states shown in Fig. \ref{projection}(a) and (b) can be connected by a second-order hopping process, with its strength denoted by $\tilde{J}$. The resulting effective Hamiltonian reads
\begin{equation} 
\hat{H}_{\textrm{eff}}=\sum_l\left[-\frac{\tilde{J}}{2}\hat{S}_{l,l+1}^+\hat{b}_l\hat{b}_{l+1}+\text{h.c.}+m\hat n_l\right],
\label{Hgauge}
\end{equation}
where $l$ labels all the even lattice sites playing a role as the matter field. The boson numbers in these sites are subjected to a constraint $n_l=b^\dagger_l b_l\leq 1$, and therefore, these bosons are considered as hard-core bosons. Here we allow certain difference between $2\delta$ and $U$, although the difference should always be small compared with $2\delta$ or $U$ such that the projection on the sub-Hilbert space is still valid. We use $m$ to denote $\delta-\frac{U}{2}$, and $m$ can be considered as the mass of a matter field particle.  

\begin{figure}[t]
\centering
\includegraphics[width=1\linewidth]{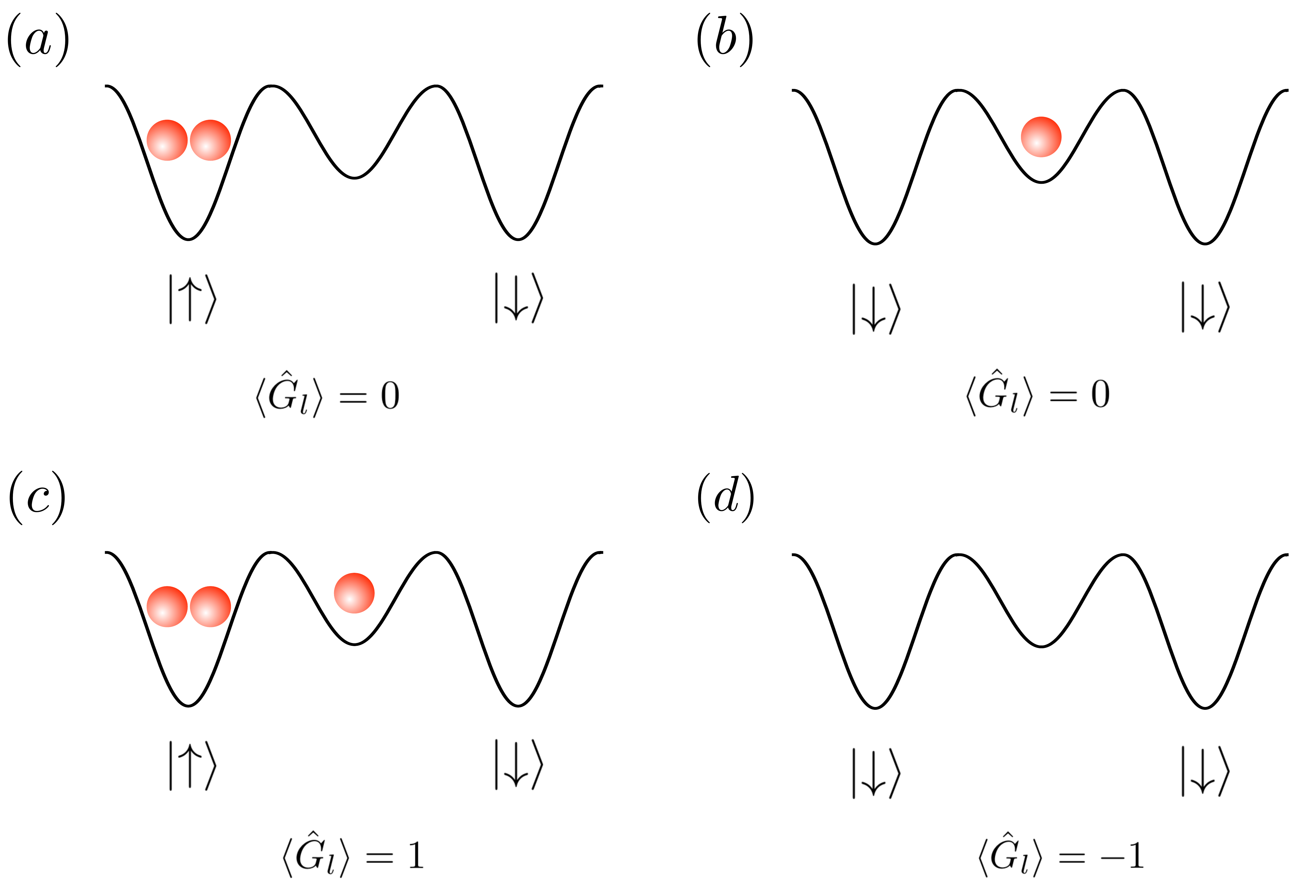}
\caption{Configurations with different gauge charge. Gauge charge $G=0$ for (a) and (b), $G=1$ for (c) and $G=-1$ for (d).}
\label{gauge_charge}
\end{figure}

\textbf{Local Gauge Symmetry.} It is easy to see that the Hamiltonian Eq. \ref{Hgauge} possesses local gauge symmetries. For any site $l$, performing a local gauge transformation $\hat{b}_l\rightarrow e^{i\theta_l}\hat{b}_l$, and simultaneously $\hat{S}^{+}_{l,l+1}\rightarrow e^{-i\theta_l}\hat{S}^{+}_{l,l+1}$, and $\hat{S}^{+}_{l-1,l}\rightarrow e^{-i\theta_l}\hat{S}^{+}_{l-1,l}$, the Hamiltonian is invariant. The local gauge symmetries force local conserved quantities. In this case, the conserved gauge charges are 
\begin{equation}
G_l=S^z_{l-1,l}+S^z_{l,l+1}+n_l, \label{eq_gauge_charge}
\end{equation} 
where $n_l$ is the particle number at site $l$ of the matter field. This can also be easily seen from the Hamiltonian Eq. \ref{Hgauge} that any spin flip increasing or decreasing $S_z$ by one is always accompanied by decreasing or increasing the boson numbers $n_l$ and $n_{l+1}$ by one. Several examples of local configurations with different $G_l$ are shown in Fig. \ref{gauge_charge}. By introducing the electric field at each link as
\begin{equation}
E_{l-1,l}=(-1)^l S^z_{l-1,l}, \label{E_field}
\end{equation}
and the physical charge at each site as $Q_l=(-1)^l n_l$, $G_l=0$ can always be rewritten as
\begin{equation}
E_{l,l+1}-E_{l-1,l}=Q_l. \label{Gauss}
\end{equation} 
This is reminiscent of the Gauss's law that ${\bf \nabla}\cdot{\bf E}=\rho$.

\begin{figure}[t]
\centering
\includegraphics[width=1\linewidth]{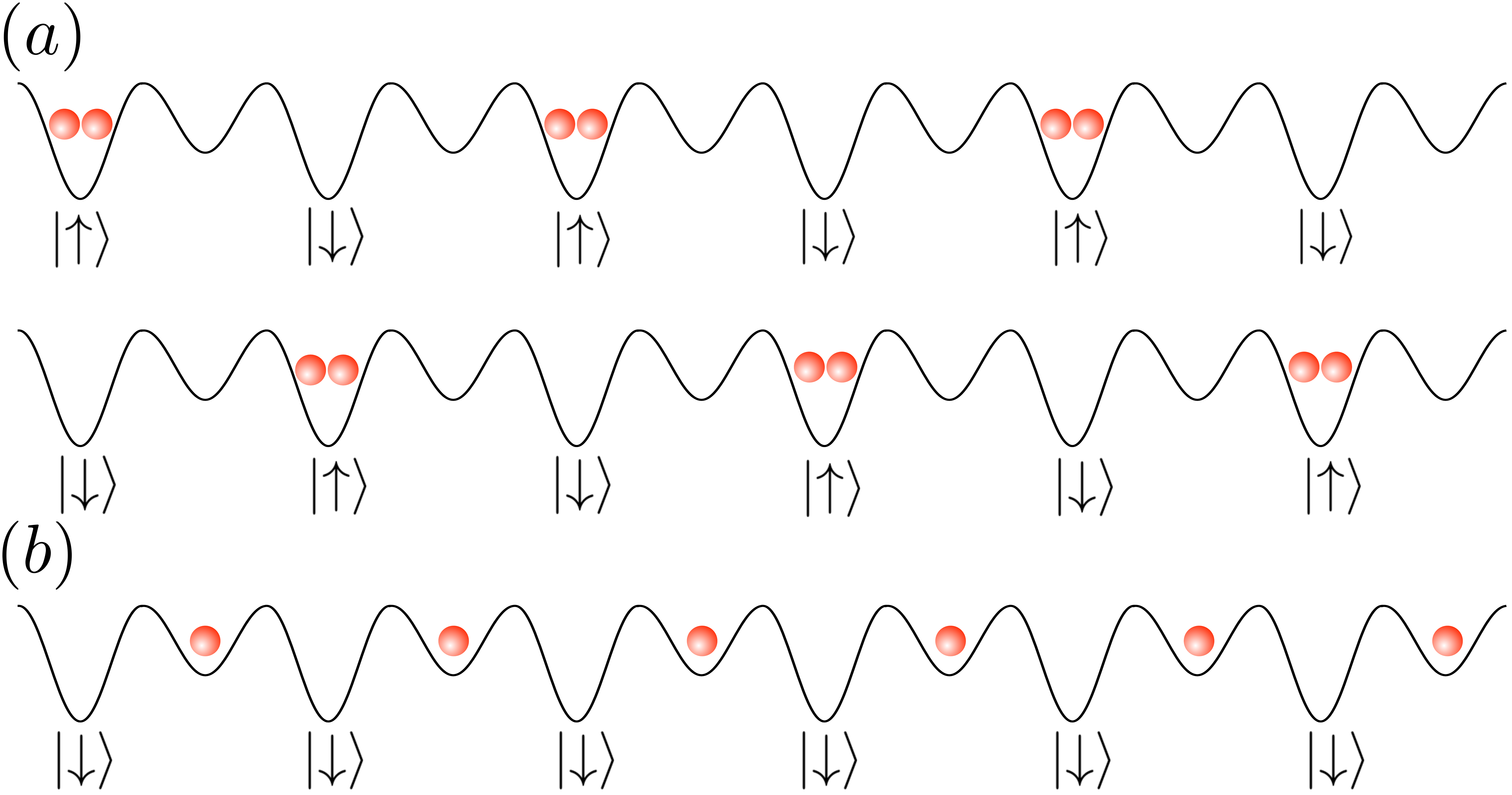}
\caption{Schematic of two phases in the $G_l=0$ gauge sector in the $m\rightarrow \infty$ limit (a) and $m\rightarrow -\infty$ limit (b).}
\label{phases}
\end{figure}

\textbf{Phase Transition.} There is a quantum phase transition in the $G_l=0$ gauge sector. It is easy to identify this transition by studying two limits of $m\rightarrow \pm \infty$ as shown in Fig. \ref{phases}. In the limit $m\rightarrow -\infty$, it favors $n_l=1$ and all matter field sites are singly occupied. Therefore, the gauge constraints $G_l=0$ force that all gauge field sites are in $\ket{\downarrow}$ states. This state respects the lattice translational symmetry. In the opposite limit $m\rightarrow \infty$, it favors $n_l=0$ and all matter field sites are empty. Therefore, the gauge constraints $G_l=0$ force $S^z_{l-1,l}+S^z_{l,l+1}=0$ and the spins in the gauge field sites form a classical anti-ferromagnetic order. Thus, this state breaks the lattice translational symmetry, and the two states shown in Fig. \ref{phases}(a) are degenerate. Therefore, we expect that there exists an Ising-type quantum phase transition as tuning $m$. In fact, the signature of this phase transition has been observed in the experiment reported in Ref. \cite{USTC}

\section{Tuning the Topological Angle} 
\begin{figure}[t]
\centering
\includegraphics[width=0.8\linewidth]{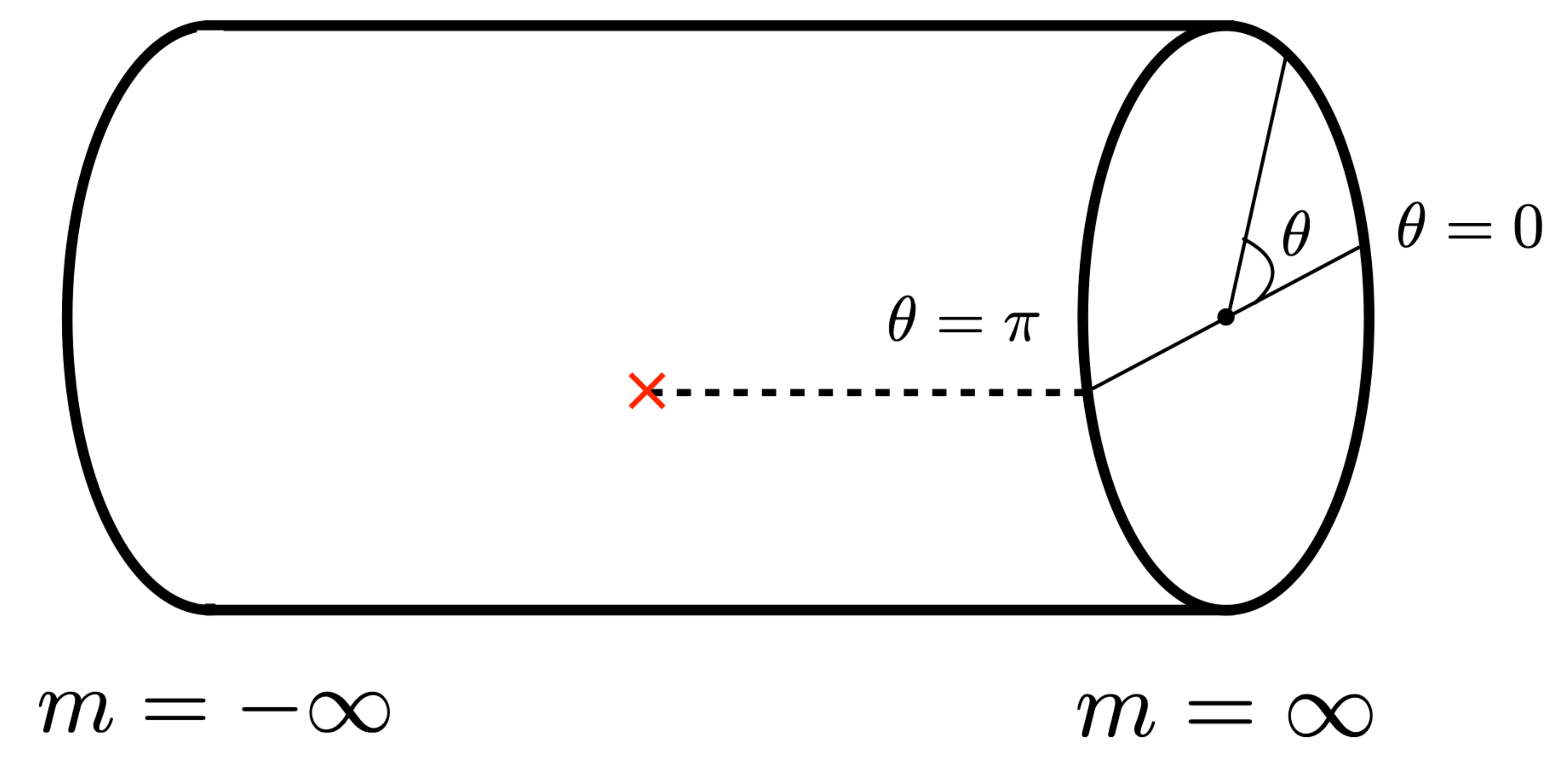}
\caption{Schematic of the phase diagram of the one-dimensional $U(1)$ LGT in term of the mass parameter $m$ and the topological angle $\theta$. The dashed line represents the deconfinement phase, and the rest of the area is the confinement phase.}
\label{gauge_LGT}
\end{figure}

A phase diagram of the $U(1)$ LGT is shown in Fig. \ref{gauge_LGT}. Aside from the mass parameter $m$, another important parameter is the topological angle $\theta$. In the $U(1)$ LGT, a conf-deconf transition only occurs when $\theta=\pi$. This transition is actually the same transition as the Ising-type transition discussed above. In other words, the situation discussed above corresponds to the $\theta=\pi$ situation in the LGT. To understand why it is the case, we need to address the following two closely related issues. Firstly, we need to introduce the parameter $\theta$ in the LGT. Secondly, the effective Hamiltonian Eq. \ref{Hgauge} only contains the gauge-matter coupling term and lacks the dynamical term of the gauge field itself. The dynamical term is an analogy of the Maxwell term in the conventional $U(1)$ gauge theory.  

For the first question, the parameter $\theta$ is introduced through the quantization value of the electric field, i.e. 
\begin{equation}
E_{l-1,l}=n-\frac{\theta}{2\pi},\  n\in\textbf{Z}.\label{eq:ThetaDef}
\end{equation}
Therefore, by the definition of Eq. \ref{E_field}, it is clear that $E_{l-1,l}$ can take values of $n-1/2$ when $\theta=\pi$. For the second question, we can add a ``Maxwell" term $gE_{l-1,l}^2$ into the Hamiltonian Eq. \ref{Hgauge}. When $g$ is large, we can only focus on $E_{l-1,l}=\pm 1/2$ when $\theta=\pi$ because other possible values of $E_{l-1,l}$ in Eq \ref{eq:ThetaDef} are not energetically favorable. Hence, Eq. \ref{eq:ThetaDef} is consistent with the definition Eq. \ref{E_field}, and the Maxwell term is a constant that can be ignored. 

With these understandings, it also becomes clear how to tune the topological angle $\theta$. We notice that Eq. \ref{E_field} can be changed to 
\begin{equation}
E_{l-1,l}=(-1)^l S^z_{l-1,l}+h. \label{E_field-1}
\end{equation}
With this new definition, the Gauss's law Eq. \ref{Gauss} still holds. However, this new definition changes the quantization value of $E_{l-1,l}$, and therefore, changes the value of $\theta$. Meanwhile, the Maxwell term now becomes
\begin{equation}
g\sum_{l}E_{l-1,l}^2=g\sum_{l}((-1)^lS^z_{l-1,l}+h)^2.
\end{equation}
Aside from constant terms, this is equivalent to adding a term $gh\sum_{l} (-1)^l S^z_{l-1,l}$ into the Hamiltonian. $h=0$ corresponds to $\theta=\pi$. Any finite $h$ tunes $\theta$ away from $\pi$, and then the conf-deconf transition no longer exists. In the original model, this corresponds to adding an alternating potential energy difference between two neighboring gauge sites. 
This alternating potential energy doubles the unit cell for the lattice translational symmetry, and the two states shown in Fig. \ref{phases}(a) are no longer related by the translational symmetry of the Hamiltonian, which is consistent with an energy splitting between them. 
Hence, from the symmetry perspective, it is also understandable that the Ising-type symmetry breaking phase transition no longer exists. 

\section{Probe Confinement and Deconfinement}

As we mentioned above, the Ising-type transition with $h=0$ is also the conf-deconf transition in the LGT with $\theta=\pi$. However, the USTC experiment has only observed evidence of anti-ferromagnetic spin order \cite{USTC} and has not observed evidence of confinement or deconfinement. Here we propose an experimental scheme to detect the confinement or deconfinement. The basic idea is to change both the gauge charge and the physical charge locally. The gauge charge cannot move due to the local gauge conservation, but the physical charge is free to move. The evidence should be:
\begin{itemize}
  \item If the physical charge is always localized around the gauge charge, this is the evidence of the confinement.
  \item  If the physical charge is delocalized and free to move, this is the evidence of the deconfinement.  
\end{itemize}

To be concrete, we first prepare the ground state $|\Psi_g\rangle$ under a fixed parameter. Then we either apply $\hat{b}^\dag_l$ to create a physical charge at site-$l$, such as the process from Fig. \ref{gauge_charge}(a) to Fig. \ref{gauge_charge}(c) that changes $G_l=0$ to $G_l=1$, or we apply $\hat{b}_l$ to annihilate a physical charge at site-$l$, such as the process from Fig. \ref{gauge_charge}(b) to Fig. \ref{gauge_charge}(d) that changes $G_l=0$ to $G_l=-1$. Experimentally, this can be achieved by the capability of single-site addressing using quantum gas microscope, together with single-site resolved density measurement and post-selection. We follow the time evolution of the quantum state $\hat{b}_l^\dag|\Psi_g\rangle$ or $\hat{b}_l |\Psi_g\rangle$ and monitor the real-space distribution of the physical charge. 

\begin{figure}[t]
\centering
\includegraphics[width=1\linewidth]{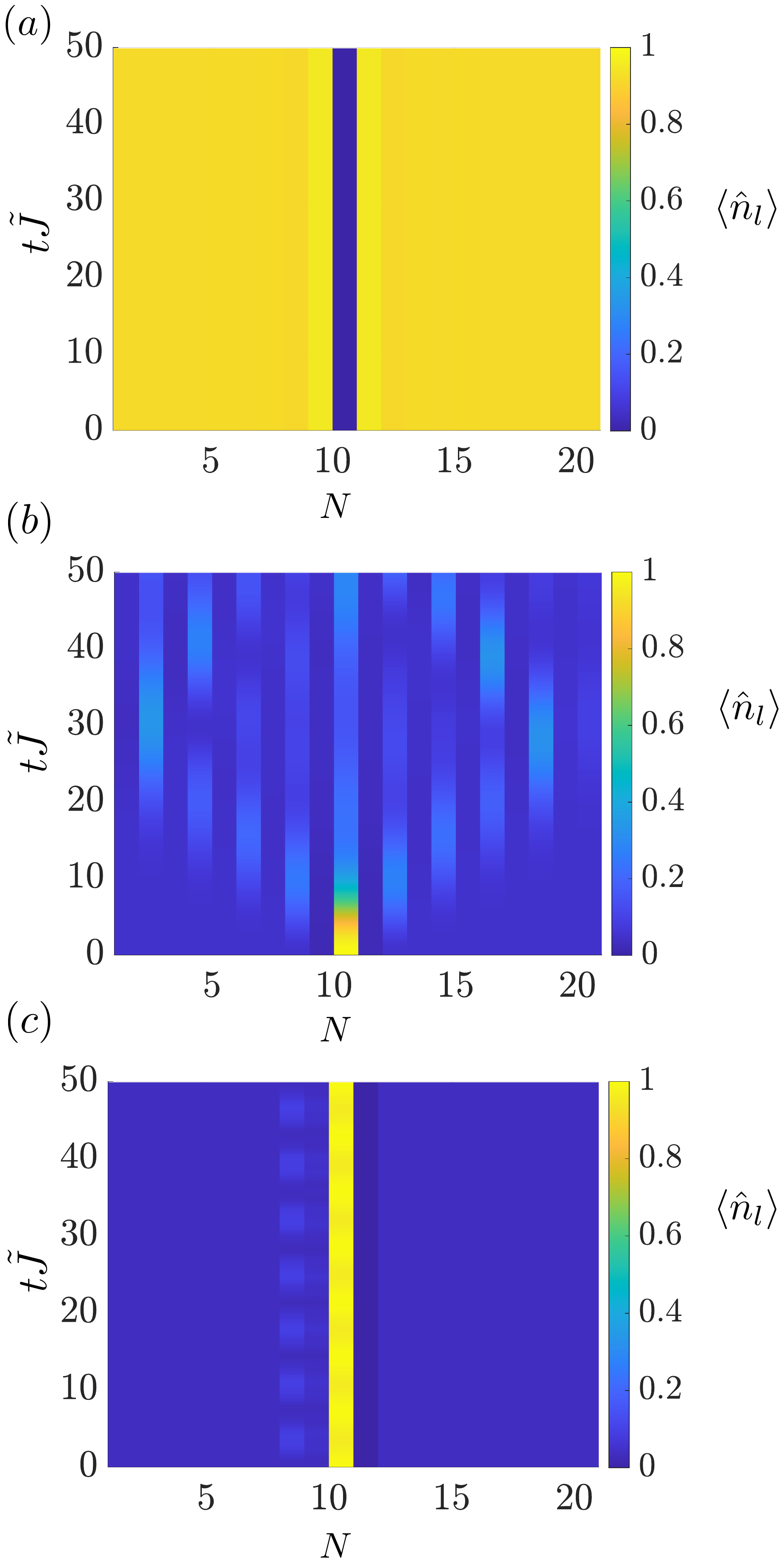}
\caption{Time evolution of the physical charge distribution after adding or removing a physical charge. (a) $m=-\tilde{J}$, $h=0$, $\theta=\pi$, confinement phase. (b) $m=\tilde{J}$, $h=0$, $\theta=\pi$, deconfinement phase. (c) $m=\tilde{J}$, $gh=0.25\tilde{J}$ and $\theta\neq \pi$.}
\label{conf-deconf}
\end{figure}

The results of numerical simulation are shown in Fig. \ref{conf-deconf} \cite{code}. In Fig. \ref{conf-deconf}(a), we show the physical charge distribution after removing a physical charge for $m=-\tilde{J}$ and $h=0$. It shows that the physical charge is always localized around the non-zero gauge charge. Instead, Fig. \ref{conf-deconf}(b) shows the charge distribution after creating a physical charge for $m=\tilde{J}$ and $h=0$. Very differently, the extra physical charge is delocalized. This difference clearly distinguishes confinement from deconfinement. Here we should note that, to obtain Fig. \ref{conf-deconf}(b), we start from a ground state that is an equal weight superposition of two symmetry breaking degenerate states. If we start from one of the symmetry breaking state, the physical charge only moves to one side, not the other side. Nevertheless, since the experiments are always averaged over two symmetry breaking states, similar results are also obtained. 

Finally, the parameters and the initial state for Fig. \ref{conf-deconf}(c) are identical to those for Fig. \ref{conf-deconf}(b), except that $gh$ is non-zero and therefore $\theta\neq \pi$. In this case, there is no phase transition and the entire parameter regime is the confinement phase. And therefore, as we show in Fig. \ref{conf-deconf}(c), the extra charge becomes localized again. 

\section{Mapping to the PXP Model}

The PXP model describes the physics of Rydberg atoms array, which reads
\begin{equation}
\hat{H}=\sum_{i}\left(-\Omega\hat{P}_{i-1}\hat{S}^x_{i}\hat{P}_{i+1}-\Delta\hat{S}^z_i\right). \label{PXP}
\end{equation}
In the model, the $\ket{\downarrow}$ denotes atoms in its ground state, and $\ket{\uparrow}$ denotes atoms in its Rydberg excited state. $\Omega$ is the coupling strength between the ground and the Rydberg states, and $\Delta$ is the effective detuning between them. Here the projection operator $\hat{P}_i$ is defined as $\hat{P}_i=(I/2-S^z_i)$, which projects the atom at site-$i$ into its ground state, and $1-\hat{P}_i$ projects the atom into its Rydberg excited state. In this case, the Rydberg blockade radius is set to be one lattice spacing, and therefore, it forbids the configuration such as $\ket{\uparrow\uparrow}$, where two neighboring atoms cannot be both in the Rydberg atom state. This constraint can also be written as $(1-\hat{P}_i)(1-\hat{P}_{i+1})|\Psi\rangle=0$ for all physical states $|\Psi\rangle$. 

Below we will show that the $U(1)$ LGT under the constraints $G_l=0$ for all $l$ is equivalent to the PXP model \cite{LGTtheory8}. The arguments follow from the following items. First, it is clear that because of the constraint Eq. \ref{eq_gauge_charge}, the occupations of matter field sites are completely determined once the gauge spin configuration is fixed. Secondly, the constraints $G_l=0$ allow $S^z_{l,l-1}+S^z_{l,l+1}=0$ or $-1$ and rule out $S^z_{l,l-1}+S^z_{l,l+1}=1$, because $n_l$ can only be zero or unity. That is to say, the constraints $G_l=0$ play the same role as the Rydberg blockade condition that forbids $\ket{\uparrow\uparrow}$ state. Thus, we can rewrite Eq.~ \ref{Hgauge} as
\begin{equation}
\begin{aligned} 
\hat{H}=\sum_l\Bigg[\hat{P}_{l-1,l}\left(-\frac{\tilde{J}}{2}\hat{S}_{l,l+1}^++\text{h.c.}\right)\hat{P}_{l+1,l+2} \\
-m(S^z_{l-1,l}+S^z_{l,l+1})\Bigg].
\label{PXP2}
\end{aligned}
\end{equation}
Here we have removed the matter field and replaced it with the projection operator to implement the requirement of the gauge constraints, and we have used $n_l=-S^z_{l,l-1}-S^z_{l+1,l}$ for $G_l=0$. Eq. \ref{PXP2} can be further simplified as
\begin{equation} 
\hat{H}=\sum_l\left[-\tilde{J}\hat{P}_{l-1,l}\hat{S}^x_{l,l+1}\hat{P}_{l+1,l+2}-2mS^z_{l,l+1}\right].
\label{PXP3}
\end{equation}
Therefore, Eq. \ref{PXP3} is exactly the same as the PXP Hamiltonian Eq. \ref{PXP} with $\Omega=\tilde{J}$ and $\Delta=2m$. 

In the PXP model, when $m$ increases toward a large positive value, it favors $S^z=1/2$ that atoms are in the Rydberg excited state. However, the Rydberg blockade forbids two neighboring atoms both in the Rydberg states. Thus, the optimum situation is half of the total atoms are excited, and the Rydberg atoms are alternatively occupied. This creates an antiferromagnetic spin configuration, and the ground state is doubly degenerate. This is the intuitive picture of how this quantum phase transition is understood in the PXP model.   

Below we will discuss the conf-deconf transition in the PXP model. There is a difference compared with what has been discussed above. Above, we have added or removed a physical charge that also changes the local gauge charge. However, the equivalence between the LGT and the PXP model requires the restriction of $G_l=0$ for all $l$. Thus, to probe the conf-deconf transition in the PXP model, we should consider a situation that does not change the gauge charge in the corresponding LGT. Thus, in the $U(1)$ LGT, we consider the physical process that flips one spin in the gauge site and simultaneously changes $n_l=0$ to $n_l=1$ (or vice versa) in its two neighboring matter sites. This prepares the state $\hat{S}_{l,l+1}^+\hat{b}_l\hat{b}_{l+1}|\Psi_g\rangle$ or $\hat{S}_{l,l+1}^-\hat{b}_l^\dagger\hat{b}_{l+1}^\dagger|\Psi_g\rangle$. Since the physical charge is defined as $Q_l=(-1)^l n_l$, changing $n_l=0$ to $n_l=1$ (or vice versa) in two neighboring matter sites corresponds to creating a pair of physical charge and anti-charge. After time evolution of the initial state, we measure the distance between two charges. The corresponding evidence should be:
\begin{itemize}
\item If two charges are always bound together, this is the evidence of the confinement phase. \item If these two charges are unbounded and free to move, this is the evidence of the deconfinement phase.
\end{itemize}

\begin{figure}[t]
\centering
\includegraphics[width=1\linewidth]{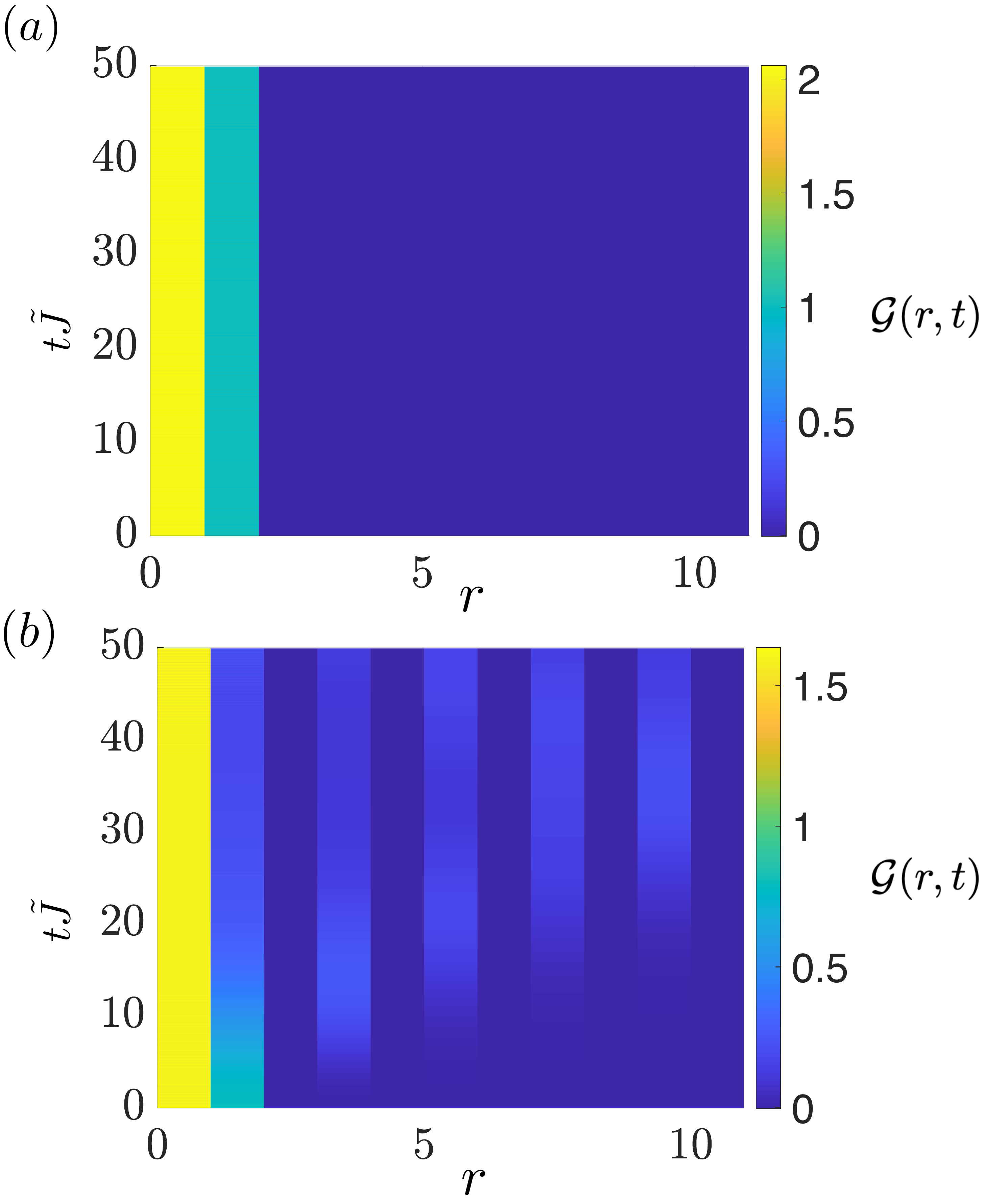}
\caption{The time evolution of $\mathcal{G}(r,t)$ as a function of $r$ and $t\tilde{J}$. $m=-10\tilde{J}$ for (a) and $m=10\tilde{J}$ for (b).}
\label{correlation}
\end{figure}

Such physics can be probed by studying the correlation function
\begin{equation}
\mathcal{G}(r,t)=\sum_l\langle\Psi(t)|(\hat{n}_l-\bar{n}_l)(\hat{n}_{l+r}-\bar{n}_{l+r})|\Psi(t)\rangle,
\end{equation}  
where $\bar{n}_l$ denotes the mean density at site-$l$ in the ground state before the spin is flipped.
$\mathcal{G}(r,t)$ is always a localized function of $r$ in the confinement phase, and is an extended function in the deconfinement phase. Using the gauge constraint $G_l=0$, we can find 
\begin{equation}
\hat{n}_l-\bar{n}_l=-\hat{S}^z_{l-1,l}-\hat{S}^z_{l,l+1}+\bar{S}^z_{l-1,l}+\bar{S}^z_{l,l+1},
\end{equation}
where $\bar{S}^z$ denotes the averaged value of $S^z$. 
Thus, we can define 
\begin{equation}
\hat{A}_{l}=-\hat{S}^z_{l-1,l}-\hat{S}^z_{l,l+1}+\bar{S}^z_{l-1,l}+\bar{S}^z_{l,l+1}.
\end{equation}
Then, in the PXP model, $\mathcal{G}(r,t)$ can be expressed as 
\begin{equation}
\mathcal{G}(r,t)=\sum_l\langle\Psi(t)|\hat{A}_l\hat{A}_{l+r}|\Psi(t)\rangle.
\end{equation}  

In Fig. \ref{correlation}, we show the time evolution of $\mathcal{G}(r,t)$ after flipping one spin in the PXP ground state. It is clear that the correlation $\mathcal{G}(r,t)$ is always localized for negative $m$ in the confinement phase, and $\mathcal{G}(r,t)$ is quickly broadened to the system size when $m$ is positive enough to enter in the deconfinement phase. 

\section{Summary}

In summary, our work proposes a protocol to directly probe confinement and deconfinement in recent experiments, and the physical systems include both the $U(1)$ LGT realized with bosons in optical lattices and the PXP model realized with the Rydberg atoms array. The LGT with gauge constraints $G_l=0$ is equivalent to the PXP model, and the local gauge constraints play the same role as the constraints from the Rydberg blockade. A quantum phase transition exists in both models as tuning the Zeeman field or the detuning. From the symmetry breaking point of view, this transition is an Ising type phase transition to antiferromagnetic states breaking the lattice translational symmetry. From the gauge theory point of view, this transition is also a conf-deconf transition. This confinement and deconfinement can be measured either by studying whether a physical charge is always localized around a gauge charge to screen it or by studying the relative motion between a physical charge and an anti-charge. The former can be directly applied to the LGT simulator with bosons in optical lattices, and the latter applies to the PXP model simulation with the Rydberg atoms array. This transition no longer exists, and the deconfinement phase disappears when an additional term is added into the Hamiltonian to explicitly break the lattice transition symmetry. This term can be interpreted as tuning the topological angle $\theta$ in the LGT. Hence, we propose that the topological angle can be tuned by applying an external symmetry breaking potential. Our proposal offers a complete route toward probing the entire phase diagram of this $U(1)$ LGT. 

\textit{Acknowledgment.} We thank Jingyuan Chen and Zhensheng Yuan for helpful discussions. The project is supported by Beijing Outstanding Young Scholar Program, NSFC Grant No.~11734010 and the XPLORER Prize. 

\textit{Note added.} For a related work, see the article submitted to arXiv on the same day by Jad C. Halimeh, Ian P. McCulloch, Bing Yang, and Philipp Hauke, entitled ``Tuning the Topological $\theta$-Angle in Cold-Atom Quantum Simulators of Gauge Theories".

\end{document}